\documentclass[aps,amsmath,amssymb,prl,twocolumn,showpacs]{revtex4}

\usepackage{graphicx}
\usepackage{times}
\usepackage{amsmath}
\usepackage{amsfonts}
\usepackage{amssymb}
\usepackage{units}
\DeclareGraphicsExtensions{.pdf,.png,.eps,.jpg}

\renewcommand{\vec}{\mathbf}

\begin{document}

\title{Engineering Quantum Spin Hall Effect in Graphene Nanoribbons via Edge Functionalization}

\author{Gabriel Aut\`es}
\author{Oleg V. Yazyev}
\affiliation{Institute of Theoretical Physics, Ecole Polytechnique F\'ed\'erale de Lausanne (EPFL), CH-1015 Lausanne, Switzerland}

\date{\today}

\begin{abstract}

Kane and Mele predicted that in presence of spin-orbit interaction graphene realizes 
the quantum spin Hall state. However, exceptionally weak intrinsic spin-orbit splitting
in graphene ($\approx 10^{-5}$~eV) inhibits experimental observation of this topological
insulating phase. To circumvent this problem, we propose a novel approach towards controlling
spin-orbit interactions in graphene by means of covalent functionalization of graphene 
edges with functional groups containing heavy elements. Proof-of-concept first-principles calculations show that very strong spin-orbit coupling 
can be induced in realistic models of narrow graphene nanoribbons with tellurium-terminated edges. We demonstrate that electronic bands with strong
Rashba splitting as well as the quantum spin Hall state spanning broad energy ranges can be realized in such systems. Our work thus opens up new horizons towards engineering topological electronic phases in nanostructures based on graphene and other materials by means of locally introduced spin-orbit interactions.
\end{abstract}


\maketitle

In their pioneering paper, Kane and Mele showed that in the 
presence of spin-orbit coupling graphene becomes a two-dimensional 
$Z_2$ topological insulator or, in other words, realizes the 
quantum spin Hall (QSH) state \cite{kan05}. This topologically 
non-trivial insulating phase is characterized by the presence of
spin-filtered edge states protected from elastic back-scattering 
and localization by time-reversal symmetry. 
While the prediction of Kane and Mele contributed tremendously 
to the development of the emerging field of topological insulators \cite{moo10,has10,qi11},
no QSH effect in graphene has been observed experimentally.
The reason being that the exceptionally weak intrinsic spin-orbit 
coupling in graphene results in a band gap of the order 
of only 10$^{-5}$~eV, according to recent theoretical 
predictions \cite{boe07,gmi09,kon10}.
   
Subsequent efforts have focused on circumventing the problem of weak
intrinsic spin-orbit coupling in graphene. 
One proposed approach for enhancing spin-orbit interactions
is based on the effect of adatoms on the electronic structure of 
graphene \cite{cas09,fab13}. More recently,  
several groups have demonstrated theoretically by depositing 
the adatoms of heavy elements with large magnitudes of atomic 
spin-orbit splitting it is possible to increase the spin-orbit gap up to tens meV \cite{abd10,wee11,jia12,zha12}. However, realizing such this strategy in practice will inevitably face the challenge of uniform 
deposition since adatoms are mobile\ cite{cha08,yaz10b} and tend to form aggregates 
due to attractive interactions. Moreover, large metal 
adatoms coverages required for enhancing spin-orbit coupling in 
graphene result in strong doping \cite{cha08,yaz10b,wee11} while covalently bonded species such 
as hydrogen adatoms give rise to undesired resonant states \cite{per06, yaz07}.

In this paper, we propose an alternative approach for engineering 
spin-orbit interactions in graphene based on covalent chemical
functionalization of the edges. The edges of graphene nanostructures 
provide a natural interface for coupling heavy-element functional 
groups to the $\pi$-electron states of graphene. By using a
modification of the Kane-Mele model we show that through introducing 
spin-orbit interaction locally, only at the edges, it is possible to realize 
the QSH state in narrow graphene nanoribbons. We then focus our
attention on realistic models of tellurium (Te) terminated graphene nanoribbons which can be 
produced using current methods of synthetic chemistry. Our first-principles
calculations show that very strong spin-orbit coupling can be induced in 
narrow Te-terminated graphene nanoribbons. In particular, we discuss
in detail two configurations that exhibit the presence of a parabolic band
with strong Rashba splitting and the robust QSH state spanning a broad 
energy range, respectively. 

\begin{figure}
\includegraphics[width=85mm]{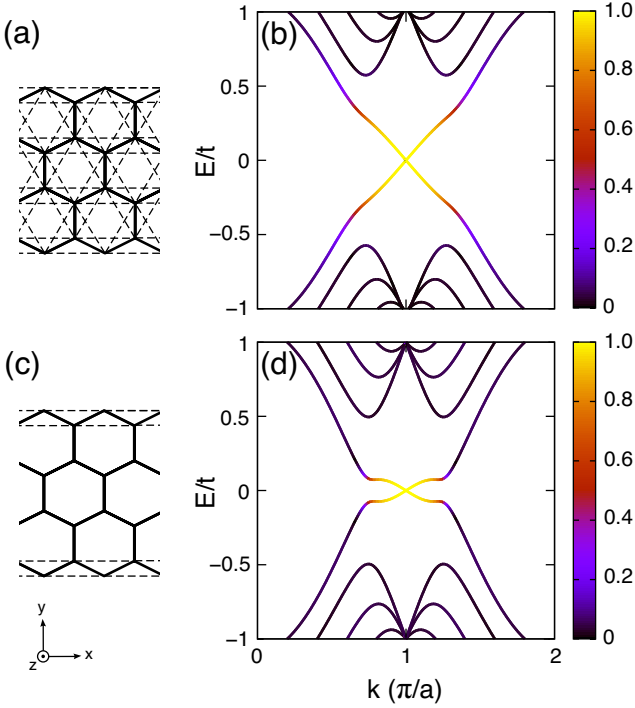}
\caption{(Color online) (a) Lattice model and (b) band structure of 
a narrow zigzag graphene nanoribbon within the Kane-Mele model at $\lambda_{\rm so}=0.1t$. 
Solid lines in the lattice model correspond to the nearest neighbor 
hoppings while dashed lines indicate all possible pairs of second-nearest-neighbor hoppings summed in the second term of Hamiltonian~(\ref{eq1}). (c) In the modified Kane-Mele model only second-nearest-neighbor hoppings involving edge atoms are taken into account
resulting in the band structure shown in (d). The bands are 
colored according to the magnitude of QSH marker ${\mathcal M}_{ik}$.
 }
\label{fig1}
\end{figure}

The band structure of graphene nanoribbons can be described accurately using the one-orbital nearest-neighbor tight-binding
model. Following Kane and Mele, it is possible to include the intrinsic spin-orbit coupling by adding a second-nearest-neighbor complex term with sign dependent on the spin of 
electron and the hopping direction \cite{kan05}.
The resulting Hamiltonian reads
\begin{equation}
{\mathcal H}=-t\sum_{\left<i,j\right>,\sigma} c_{i\sigma}^{\dagger}c_{j\sigma} 
+ i\lambda_{\rm so}\sum_{\left<\left<i,j\right>\right>, \sigma}  \nu_{ij} c_{i\sigma}^{\dagger} \sigma_z c_{j\sigma},
\label{eq1}
\end{equation}
where $\left<i,j\right>$ and  $\left<\left<i,j\right>\right>$ indicate the pairs of first and second nearest neighbors [Fig.~\ref{fig1}a], respectively, and $\sigma$ is the spin index. Parameter $t \approx 2.7$~eV is the tight-binding hopping energy \cite{cas05} 
while $\lambda_{\rm so}$ defines the strength of spin-orbit coupling. $\nu_{ij}=\pm 1$ is the site-dependent Haldane factor given by $\nu_{ij}=(\vec{d}_{ik} \times \vec{d}_{jk})/|\vec{d}_{ik} \times \vec{d}_{jk}|$ with $\vec{d}_{ik}$ and $\vec{d}_{jk}$ being the vectors connecting second nearest neighbors $i$ and $j$ with their common neighbor $k$.\cite{hal88} $\sigma_z$ is the corresponding Pauli matrix describing the spin of the electron.

The introduction of the spin-orbit term opens a band gap 
$\Delta_{\textrm{so}}=6\sqrt{3}\lambda_{\textrm{so}}$ 
at the Dirac
points of otherwise semimetallic graphene making it a two-dimensional $Z_2$ topological insulator. When one-dimensional nanostructures of 
graphene are considered, spin-orbit coupling lifts the degeneracy of
zero-energy edge states \cite{nak96,ryu02,yaz13} resulting in a crossing of linear dispersion bands as
illustrated in Fig.~\ref{fig1}b for the case of a narrow graphene 
strip (nanoribbon). At this band crossing the electronic states 
of opposite spins are localized at the opposite edges of the 
nanoribbon while time-reversal symmetry is preserved. In other words,
the nanoribbon is in the QSH state.

Let us now consider a situation in which spin-orbit interaction are 
introduced only at the edges due to, for instance, the presence of 
heavy-element functional groups terminating the edge. One can expect
that such local spin-orbit coupling will have a significant effect
on the $\pi$-electron states in graphene due to finite size effects 
and the fact that low-energy states in graphene nanostructures are
strongly localized at the edges \cite{nak96,wim10}. In order to model this situation,
we restrict summation in the second term of Hamiltonian~(\ref{eq1}) only
to the pairs of second nearest neighbor atoms located at the edge
as shown in Fig.~\ref{fig1}c. As in the case of the standard Kane-Mele model,
the resulting band structure calculated using the same magnitude of 
$\lambda_{\rm so}=0.1t$ exhibits a band crossing at $k=\pi/a$ [Fig.~\ref{fig1}d].

\begin{figure}
\includegraphics[width=85mm]{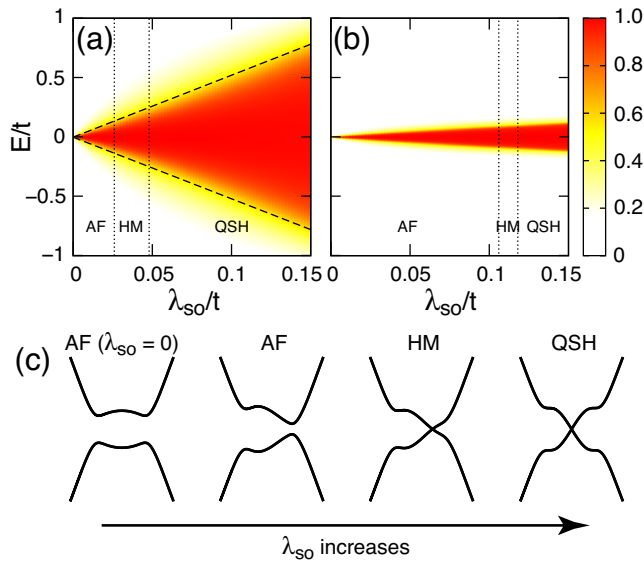}
\caption{(Color online) Magnitude of the QSH marker ${\mathcal M}_{ik}$ as a function of energy $E$ and spin-orbit coupling
strength $\lambda_{\rm so}$ for a zigzag nanoribbon in the Kane-Mele model with (a) a full spin-orbit coupling term and (b) spin-orbit interactions introduced only at the edges. 
The dashed lines in (a) indicate the bulk spin-orbit gap $\Delta_{\textrm{so}}$. The dotted lines delimit the regions of
existence of electronic phases driven by electron-electron interactions. (c) Schematic drawing of the evolution of edge-state
band dispersion upon increasing $\lambda_{\rm so}$ in the presence
of electron-electron interactions.  }
\label{fig2}
\end{figure}

In order to identify the presence of the QSH state we propose a simple marker function which signalizes the spin-filtered character of electronic states. 
For state $i$ of momentum $k$, we calculate the vectorial quantity with
components
\begin{equation}
\label{indicator}
{\mathcal M}^\alpha_{ik} = \frac{2}{W} \int \psi^\dagger_{ik}(\mathbf{r})  y \sigma_\alpha \psi_{ik}(\mathbf{r}) d\mathbf{r},
\end{equation}
where $W$ is the width of the nanoribbon and $\sigma_\alpha$ are the Pauli matrices ($\alpha = x,y,z$). By convention, we assume that 
the $y$ direction is transverse to the nanoribbon and lies in its plane; $y=0$ is located in the middle of the nanoribbon.
Ordinary  spin-degenerate states are characterized by $|{\mathcal M}_{ik}| = 0$ while the maximum magnitude $|{\mathcal M}_{ik}| = 1$ corresponds to spin-filtered states completely localized on the edge atoms. In the presence of time-reversal symmetry, a non-zero magnitude of ${\mathcal M}_{ik}$ indicates the QSH phase while its
direction corresponds to the spin direction of the edge state.
In the case of Kane-Mele model with all second-nearest-neighbor atoms
summed, the edge states are fully spin-filtered in a broad energy 
range [Fig.~\ref{fig1}b]. Importantly, the model with spin-orbit interactions introduced at the edges only also results in 
complete spin filtering, but the energy range in which the 
system is in the QSH regime is reduced, assuming the same value of 
$\lambda_{\rm so} = 0.1t$. To gain a more quantitative
insight we analyze the magnitude of ${\mathcal M}_{ik}$ as a function 
of energy $E$ and spin-orbit coupling strength $\lambda_{\rm so}$
for both models [Figs.~\ref{fig2}a,b]. In the case of the standard Kane-Mele model the QSH regime spans the spin-orbit band gap $\Delta_{\textrm{so}}=6\sqrt{3}\lambda_{\textrm{so}}$ [dashed lines
in Fig.~\ref{fig2}a]. For the modified Kane-Mele model we also find 
a linear dependence of the QSH energy range on $\lambda_{\rm so}$,
although scaled down by approximately a factor of 5 compared to the former case. In both cases, the direction of the edge-state electron spins is along the $z$ axis by construction [Eq.~(\ref{eq1})]. 
Another important difference between the two models is the width dependence of the results. In the case of standard Kane-Mele model the QSH energy range is independent of nanoribbon width. However, it 
decays with increasing the nanoribbon width for the modified Kane-Mele model due to the decrease of the energy gap between bulk-like valence and conduction bands.

We note that the Rashba spin-orbit term has been omitted in our model. It was demonstrated by Kane and Mele that while the Rashba term violates the conservation of $S_z$, the QHS phase is present for the values of the Rashba parameter $\lambda_{\textrm{R}} < 2\sqrt{3}\lambda_{\textrm{so}}$ \cite{kan05-2}.
Similarly, for the modified edge-only Kane-Mele model our calculations show that the QSH state persists for $\lambda_{\textrm{R}} \lesssim \lambda_{\textrm{so}}$. 

We would now like to discuss the possible competing role of 
electron-electron interactions in graphene systems with locally
induced spin-orbit interactions. For the purpose of our discussion 
we introduce into our Hamiltonian the Hubbard term of the form
\begin{equation}
{\mathcal H}'=U\sum_{i} n_{i\uparrow}n_{i\downarrow},
\label{eq3}
\end{equation}  
where  $n_{i\sigma}=c_{i\sigma}^{\dagger}c_{i\sigma}$ is the spin-resolved electron density on site $i$ and $U>0$ defines the magnitude of the on-site Coulomb repulsion. For practical reasons, we evaluate this term using the mean-field approximation
\begin{equation}
{\mathcal H}'_{\rm mf}=U\sum_{i} \left( n_{i\uparrow}\left <n_{i\downarrow} \right > + n_{i\downarrow}\left <n_{i\uparrow}\right> - \left<n_{i\downarrow}\right>\left <n_{i\uparrow}\right> \right).
\label{eq4}
\end{equation}
In the absence of spin-orbit coupling the introduction of 
electron-electron interactions results in a magnetically ordered state characterized by ferromagnetic correlations along the edges and  
antiferromagnetic correlations across the nanoribbon \cite{fuj96,son06}. The state is gapped as shown in the 
schematically depicted band structure [AF ($\lambda_{\rm so} = 0$)
in Fig.~\ref{fig2}c] and breaks time-reversal symmetry.
Spin-orbit interactions break the valley symmetry [AF in Fig.~\ref{fig2}c] and result in band gap closing above some critical
spin-orbit interaction strength $\lambda^{c1}_{\rm so}$ \cite{sor10}.
The new phase is a valley half-metal [HM in Fig.~\ref{fig2}c].
Further increase of the spin-orbit interaction strength above
$\lambda^{c2}_{\rm so}$ suppresses the magnetic ordering, thus
reestablishing the QHS state and time-reversal symmetry [QSH in Fig.~\ref{fig2}c]. For the two discussed models we 
calculate the critical values of $\lambda_{\rm so}$ assuming $U/t=1$
consistent with the results of first-principles calculations and 
several experimental investigations \cite{yaz08}. In the case of full
Kane-Mele model treatment [Fig.~\ref{fig1}a] we find $\lambda^{c1}_{\rm so} = 0.026t$ and $\lambda^{c2}_{\rm so} = 0.048t$. Restricting the range of spin-orbit interactions to edge
atoms only [Fig.~\ref{fig1}c] extends the domain of existence of the
AF phase to $\lambda^{c1}_{\rm so} = 0.106t$ while the QSH phase 
emerges at $\lambda^{c2}_{\rm so} = 0.118t$ ($\approx 0.3$~eV). It 
is worth noting that the latter value is well below the atomic 
spin-orbit splittings in many heavy elements (e.g. 0.49~eV in Te 
and 1.25~eV in Bi \cite{wit74}), thus confirming the possibility of 
realizing the QSH phase in edge-functionalized narrow graphene
nanoribbons, even in the presence of competing electron-electron
interactions. Moreover, the interplay between the effects of 
spin-orbit and electron-electron interactions can 
be controlled to some degree by changing crystallographic 
orientation of the edges \cite{yaz11,aut12}.

\begin{figure*}
\includegraphics[width=170mm]{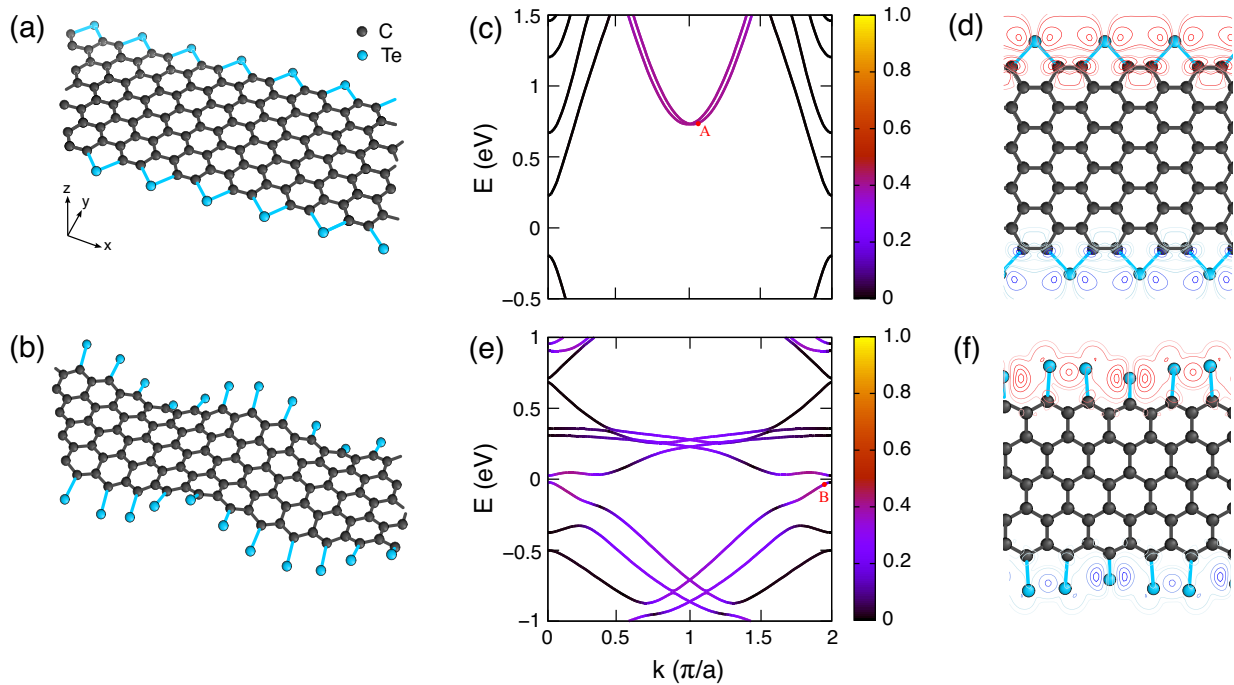}
\caption{(Color online) Relaxed atomic structures of Te-terminated (a) 
armchair and (b) zigzag graphene nanoribbons. (c) Electronic band 
structure of the Te-terminated armchair graphene nanoribbon with color-coding
indicating the magnitude of ${\mathcal M}_{ik}$. The energy is given 
relative to the Fermi level. (d) Contour plot of the spatial distribution of 
spin-filtered edge states labeled A in the band structure in panel (c). 
The spin-filtered states with opposite spin directions are distinguished 
by color. (e) Band structure and (f) spatial distribution of 
spin-filtered edge states (labeled B) in the Te-terminated zigzag graphene nanoribbon. }
\label{fig3}
\end{figure*}

In order to provide a proof of concept of the presented model we
performed a detailed first-principles investigation of realistic models of graphene nanoribbons with 
edges terminated by functional groups containing heavy elements.
Our first-principles calculations were performed within the density functional theory framework employing the local density approximation (LDA) as implemented in the QUANTUM-ESPRESSO package \cite{gia09}. Spin-orbit effects were accounted for using fully relativistic
pseudopotentials acting on valence electron wavefunctions represented in the two-component spinor form \cite{dal05}. A plane-wave cutoff
of 30~Ry was used for the wave-functions. Atomic positions were fully relaxed. 
Below, we focus on two representative examples involving 
tellurium (Te), a heavy element with large atomic spin-orbit 
splitting \cite{wit74} and chemical properties very similar
to that of isoelectronic sulfur for which synthetic organic 
chemistry is well 
established. The atomic configurations of the proposed model systems 
are motivated by the recent advances in producing atomically-precise
graphene nanoribbons as discussed below. 
The first configuration represents a 1-nm-wide armchair
graphene nanoribbon with edge termination composed of very stable
tellurophene fragments. Such narrow armchair nanoribbons can be 
produced with atomic precision using the recently developed chemical
self-assembly route \cite{cai10}. Fairly complex nanostructures based
on thiophene units (the sulfur analogue of tellurophene) can be 
synthesized using standard organic chemistry methods (for examples,
see Ref.~\onlinecite{che06}). The second configuration is a zigzag
graphene nanoribbon with a larger Te coverage density at the edges. 
The sulphur analogues of such nanostructures have been produced
recently by fusing sulfur-rich precursor molecules inside a carbon 
nanotube matrix through heating or electron beam irradiation \cite{chu11,cha12}.
In order to account for the experimentally 
observed out-of-plane deformation resulting from steric repulsion of
Te atoms at the edge we consider a supercell model composed of 
6 unit cells.

Figure~\ref{fig3}c shows the band structure of Te-terminated 
armchair nanoribbon [Fig.~\ref{fig3}a]. The parabolic valence 
and conduction 
bands with extrema at $k=0$ and the presence of a direct band 
gap of 0.42~eV are typical for 
$\pi$-symmetry states subjected to quantum confinement in armchair graphene
nanoribbons \cite{nak96}. However, we note the presence of an additional band 
at $\approx$0.5~eV above the conduction band maximum which displays
a clear Rashba-type splitting ($\alpha_{\rm R} =0.014$~eV~\AA) and 
a large magnitude of ${\mathcal M}_{ik}$. Analysis of the spatial 
distribution of the Rashba-split states [Fig.~\ref{fig3}d for the momentum and the
energy indicated by label A in Fig.~\ref{fig3}c] shows that they are formed
by the hybridization of edge C and Te orbitals. These electronic states
are spin degenerate but localized at the opposite edges (that is, 
spin-filtered); the spin is oriented along the $y$ direction. 
Strictly speaking, such a system is not in the QSH 
regime, but this example proves that the edge functionalization of 
graphene nanoribbons can result in strong spin-orbit driven effects, 
in particular, the emergence of spin-filtered edge states.

We will now discuss the second model characterized by a significantly 
larger coverage density of Te atoms on the edges. In this case,
hybridization of graphene $\pi$-states with Te orbitals strongly modifies
the band-structure features that are typical for zigzag graphene 
nanoribbons passivated by ``neutral'' functional groups such as
hydrogen atoms. However, we note the dominance of low-dispersion 
bands close to the Fermi level ($E=0$) remnant of the edge-state 
flat band in H-terminated graphene nanoribbons. As in the 
previous model, the ground state configuration of the Te-terminated
zigzag nanoribbons shows no magnetic ordering, hence  
time-reversal symmetry is preserved. The spatial orientation of the spin is also along the $y$ direction. Importantly, the electronic
bands at low energies form a number of crossings at the Kramer's 
degeneracy points $k=0$ and $k=\pi/a$ characterized by large 
magnitudes of ${\mathcal M}_{ik}$. 
One  such crossing, even though characterized by a band gap opening 
of 47~meV, is situated at the Fermi level of the charge-neutral
graphene nanoribbon [Fig.~\ref{fig3}e]. We ascribe the gap opening
to the hybridization between the edge states localized at the 
opposite edge of the nanoribbon analogous to the band gaps  observed in thin films of bulk topological insulators \cite{liu10,yaz10,zha10}.
Importantly, there are not 
other bands present in the energy range $-$0.32~eV~$< E <$~0.26~eV. 
The corresponding states are spin-filtered and the system in the 
QSH regime in this energy range. The relevant
electronic states are mostly Te-derived as shown in Fig.~\ref{fig3}f.


To summarize, we proposed a new method for engineering spin-orbit 
interactions in graphene by means of the covalent chemical functionalization
of the edges. First-principles calculations performed on realistic models 
of tellurium-terminated graphene nanoribbons demonstrate the efficiency 
of our approach and the possibility of realizing the quantum spin Hall 
state spanning broad energy ranges. Our work thus opens up new perspectives 
for controlling spin-orbit interactions as well as the practical realization 
of novel topological electronic phases in graphene nanostructures by means 
of bottom-up chemical routes.

We would like to thank M. Franz and Z. Zhu for discussions. This work was supported by the Swiss National Science Foundation (grant No. PP00P2\_133552) and by the Swiss National Supercomputing Centre (CSCS) under projects s336 and s443.

\end{document}